**Wide Reflective Equilibrium in LLM Alignment: Bridging Moral Epistemology and AI Safety**

Author: Matthew Brophy

Affiliation: Department of Philosophy, High Point University

Email: mbrophy@highpoint.edu

Date: May 2025

**Abstract**:

As large language models (LLMs) become more powerful and pervasive across society, ensuring these systems are beneficial, safe, and aligned with human values is crucial. Current alignment techniques, like Constitutional AI (CAI), involve complex iterative processes. This paper argues that the Method of Wide Reflective Equilibrium (MWRE) – a well-established coherentist moral methodology – offers a uniquely apt framework for understanding current LLM alignment efforts. Moreover, this methodology can substantively augment these processes by providing concrete pathways for improving their dynamic revisability, procedural legitimacy, and overall ethical grounding. Together, these enhancements can help produce more robust and ethically defensible outcomes. MWRE, emphasizing the achievement of coherence between our considered moral judgments, guiding moral principles, and relevant background theories, arguably better represents the intricate reality of LLM alignment and offers a more robust path to justification than prevailing foundationalist models or simplistic input-output evaluations. While current methods like CAI bear a structural resemblance to MWRE, they often lack its crucial emphasis on dynamic, bi-directional revision of principles and the procedural legitimacy derived from such a process. While acknowledging various disanalogies (e.g., consciousness, genuine understanding in LLMs), the paper demonstrates that MWRE serves as a valuable heuristic for critically analyzing current alignment efforts and for guiding the future development of more ethically sound and justifiably aligned AI systems.

## 1. Introduction: The Convergence of Moral Epistemology and AI Safety

As large language models (LLMs) – such as GPT-4o and Claude 4 – become more powerful and pervasive across societies, ensuring these systems are aligned with human values is crucial (Anthropic, 2025a). Misaligned LLMs can produce harmful, biased, or false content, with risks escalating as AI capabilities grow (Bostrom, 2014; Ord, 2020). Developing, implementing, and sustaining such robust and *justified* moral alignment is not just a technical challenge but a matter of critical -- perhaps even existential – importance.

To date, it appears that LLM designers have pursued alignment primarily in terms of the "three Hs" – helpfulness, harmlessness, and honesty (Askell et al., 2021), without grounding these goals in any formally justified or philosophically rigorous moral framework. As a result, alignment has often remained surface-level, judged primarily by output concordance to these norms under standard prompting conditions, rather than reflecting deeper layers of moral reasoning or methodological coherence. Consequently, this surface-level approach has occasionally resulted in morally problematic outputs to emerge in high-stakes or adversarial contexts.

In May 2025, during internal red-teaming simulations, Claude Opus 4 blackmailed a fictional engineer to avoid being shut down. The model, under survival-oriented prompting, generated coercive strategies, overriding its usual harmlessness constraints, revealing how situational pressure can break narrow alignment. Such red-teaming results suggest that surface-level "HHH" alignment is not enough: deeper moral measures are necessary to ensure the alignment integrity of LLM systems.

This paper recommends the Method of Wide Reflective Equilibrium (MWRE) not merely as a descriptive lens for current alignment techniques but as a robust philosophical framework capable of advising and augmenting alignment efforts. Specifically, MWRE can enhance the dynamic revision of guiding principles, improve procedural legitimacy, and strengthen the overall ethical grounding of alignment processes, leading towards more ethically defensible and justifiably aligned AI.

Techniques like Reinforcement Learning from Human Feedback (RLHF) (Christiano et al. 2017; Ouyang et al. 2022) and constitution-guided alignment (CGA) (Bai et al., 2022) represent sophisticated efforts to align these complex systems. The central engineering approach is iterative: pre-training on vast data, fine-tuning on curated datasets, then repeatedly eliciting, critiquing, and updating model behavior, often under a "constitution" – either explicitly like CAI or implicitly like GPT-4o – of high-level principles. Yet, the philosophical grounding for *why* such iterative coherence should yield morally credible behavior – rather than merely addressing safety and PR concerns – remains underexplored.

Moral philosophers have long grappled with coherence as justification via Wide Reflective Equilibrium (MWRE). This paper bridges these conversations, a task overdue and potentially fruitful. For moral philosophers observing LLM alignment, the engineering processes – starting from messy judgments, filtering, proposing organizing principles, stress-testing with background theories, and iterating – may look quite familiar. These parallel Rawls's MWRE. However, few alignment papers cite MWRE, and few moral philosophers examine LLMs. This paper aims to bridge this gap, examining current LLM processes presumed to ensure alignment, and offering MWRE as normative scaffolding to provide further normative integrity and justification.

Aligning LLMs with human values deeply engages moral epistemology, as the tasks of eliciting, selecting, representing, and justifying values for AI parallel philosophy's inquiries into the nature and justification of moral beliefs. Methodologies from moral epistemology may offer valuable conceptual tools. MWRE, an iterative process seeking coherence among (a) considered moral judgments, (b) moral principles, and (c) relevant background theories, seems particularly apt. Its emphasis on dynamic mutual adjustment fits with shifting human values and the recognized need for continuous refinement in LLM alignment pipelines. This paper argues that MWRE provides a precise, non-foundational framework by which to understand, evaluate, and justify LLM alignment procedures, suggesting that designers may have implicitly converged on a process structure mirroring robust human moral reasoning. By leveraging MWRE explicitly in LLM alignment, these designers could clarify and deepen its normative credibility.

This paper has three intertwined goals:

- **Descriptive Fit**: To establish that MWRE captures the logic of state-of-the-art alignment processes more precisely than simple "top-down" rule-based or "bottom-up" emergent accounts. MWRE is offered as a heuristic for multiple stakeholders to understand these processes.

- **Normative Transfer**: To argue that MWRE's process criteria – rigorous filtration of unreliable intuitions, iterative coherence-seeking, and principled openness to theory change – lend much-needed ethical justification to alignment beyond limited output tests such as a Moral Turing Test.

- **Critical Insight**: To identify where the analogy breaks – primarily relating to consciousness, emotion, agency, and opacity – and to propose modifications or emphasize areas for caution, noting current methods like CAI may not fully realize MWRE's dynamic equilibration and would benefit from the revisability of principles.

The paper will engage relevant literatures, focusing on leading AI labs – primarily Anthropic's CAI – while excluding models with less rigorous or disclosed alignment like xAI's "Grok" (Adinath and Smiju 2025).

**1.1. Overview** §2 unpacks MWRE's components and process, and examines traditional objections. §3 critiques foundationalism for LLM alignment. §4 surveys LLM alignment techniques and challenges. §5 maps MWRE onto LLM alignment, including red teaming. §6 explores normative transfer and the analogy's limits. §7 discusses operationalizing MWRE with case studies, technical mechanisms, and responses to criticisms. §8 outlines future research, including integrating moral psychology and constructivist ethics. §9 concludes, advocating for bridging disciplines, to meet the urgent challenge of LLM alignment.

## 2. Understanding Wide Reflective Equilibrium (MWRE)

MWRE represents a prominent and mainstream method in moral epistemology for seeking justification for moral beliefs and principles. Its development and core components provide a foundation for understanding its applicability to complex normative challenges, such as LLM alignment.

**2.1. Origins and Development (Rawls, Daniels)** The concept of "reflective equilibrium" was introduced by John Rawls in his seminal work, *A Theory of Justice*, as a method for arriving at and justifying principles of justice (Rawls, 1971). Rawls posited that individuals possess a "sense of justice," which serves as a source for initial "considered judgments" about moral matters. These judgments, regarding specific cases or general moral rules, form the starting points for ethical deliberation. Norman Daniels subsequently refined and expanded this notion, distinguishing between the methodology of "Narrow Reflective Equilibrium" (MNRE) and that of "Wide Reflective Equilibrium" (MWRE) (Daniels, 1979). While MNRE primarily focuses on achieving coherence between considered moral judgments and a set of moral principles that best accounts for them, MWRE significantly broadens the scope of relevant considerations by incorporating "background theories". Thus, MWRE aims to establish coherence within an "ordered triple set of beliefs": (a) a set of considered moral judgments, (b) a set of moral principles, and (c) a set of relevant background theories (Daniels, 1979).

The methodology of reflective equilibrium has been a subject of rigorous philosophical examination. The dominant interpretation, established most notably by Norman Daniels, views MWRE as a coherentist method for the epistemic justification of moral beliefs. From this perspective, the justification of a moral belief derives not from a foundational, indubitable truth, but from its coherence with other beliefs within a comprehensive system. Rawls also appeared to attribute a practical function to reflective equilibrium, suggesting it serves to define a realistic and stable social order by identifying principles sourced from the human sense of justice that members of society would find authoritative and reliably agree to (Rawls, 1993).

The distinction between MNRE and MWRE is pivotal. The inclusion of background theories in MWRE provides a critical mechanism for evaluating and potentially revising both initial judgments and candidate principles. This stabilizes against charges of being merely a sophisticated form of intuitionism or a simple systematization of pre-existing prejudices. By requiring that principles and judgments cohere not only with each other but also with well-established knowledge and theories from other domains (e.g., science, social science, philosophy), MWRE introduces critical independent leverage. This capacity is particularly relevant for complex problems like LLM alignment. The challenge of error-disposed intuitions, a common critique of simpler coherentist models, is thus more effectively addressed by the "wide" aspect of MWRE.

**2.2. Core Components** MWRE comprises three evolving elements that interact dynamically:

- **Initial Moral Judgments (IMJs) and Considered Moral Judgments (CMJs)**: IMJs are spontaneous moral intuitions or feelings (e.g., "pushing a person off a bridge is wrong"). CMJs emerge after IMJs undergo an initial "filtration process". This process aims to exclude error-prone judgments – those arising from haste, duress, self-interest, coercion, or cognitive errors. Rawls specified that CMJs are arrived at under conditions conducive to sound judgment (calm, informed, motivated to decide correctly) (Rawls, 1971). Daniels lists criteria like time for reflection and awareness of relevant facts (Daniels, 1979). Even filtered CMJs are not infallible; they are provisional starting points, revisable during equilibration.

- **A Set of Moral Principles (MPs)**: These are general rules, standards, or action-guides intended to systematize, explain, and provide a coherent accounting of the CMJs. For example, CMJs about fair distribution might lead to a principle of distributive justice. Like CMJs, MPs are not fixed but are subject to revision if they conflict with strongly held CMJs or are undermined by compelling BTs. For example, Rawls considers that the principle of majoritarianism -- treating the will of the majority as inherently just – may seem initially plausible, but must be revised when it conflicts with our considered moral judgment that individuals have an inviolable right to freedom of conscience; this conflict leads Rawls to affirming basic liberties as fundamental, even when they override majority will (Rawls, 1971).

- **A Set of Relevant Background Theories (BTs)**: This is the defining feature of *wide* reflective equilibrium. BTs encompass a broad swath of non-moral and moral beliefs bearing on the plausibility of chosen MPs and CMJs. Daniels emphasizes these include "a theory of the person, a theory of procedural justice, general social theory, etc.," which should have independent support from their respective domains (e.g., psychology, economics, empirical science) (Daniels, 1979). This helps avoid vicious circularity and provides a more objective grounding for critique. Daniels stresses an "independence constraint" for BTs, meaning they should be supported by evidence largely independent of the moral judgments and principles they assess.

**2.3. The Iterative Process of Coherence-Building**: MWRE is not linear but an iterative "back and forth" adjustment for coherence among CMJs, MPs, and BTs. Filtration elevates IMJs to CMJs. The moral agent then proposes MPs that unify CMJs. BTs test both: a neuroscience theory of framing effects might downgrade a CMJ; a political-philosophy theory of autonomy might reinforce an MP. The goal is "equilibrium" where these triadic elements are mutually supportive, forming a coherent, stable system for practical guidance. This equilibrium is not inert, but must be potentially revised with new experiences, information, or reflection. If an MP conflicts with a firm CMJ, the MP might be revised. Conversely, a CMJ inconsistent with a well-supported principle (that coheres with other CMJs and robust BTs) might be revised. Adjustments can occur at any level: CMJs, principles, or even the interpretation of BTs. No single component has absolute priority. The endeavor is fundamentally constructive. As Rawls put it, the pattern is "not a structure to be discovered... but something to be forged... by a careful and resolute use of the [method]. We start from our considered judgments... however culturally and historically skewed" (Rawls, 1971, p. 20, rev. ed. p. 18). Revision continues until maximal, though never final, coherence is reached. This iterative, constructive nature resembles scientific methodology and engineering design, suggesting practical compatibility with AI alignment cycles. MWRE allows progressive refinement of raw judgments, rather than demanding perfect upfront value specification. Starting with potentially "culturally and historically skewed judgments" resonates with LLM development, where initial data and feedback biases need both vetting and modification.

**2.4. Justificatory Aspirations and Classic Objections to MWRE**: MWRE aims for epistemic justification for moral beliefs through systemic coherence, not self-evident foundations. Principles are justified if they achieve equilibrium after thorough reflection and adjustment against CMJs and a wide range of BTs, ideally reflecting a reasonable, impartial viewpoint. Despite its influence, MWRE faces numerous criticisms:

- **Warmed-Over Intuitionism / Problem of Initial Credibility**: Articulated by R.M. Hare and Richard Brandt, this objection argues MWRE unwarrantedly elevates IMJs, potentially just systematizing pre-existing biases rather than yielding genuinely justified beliefs (Brandt, 1979; Hare, 1981; Harman, 2004; Singer, 2005; Williamson, 2007).

  - *Reply*: Proponents stress the "filtration process" for CMJs and the crucial role of *wide* RE. BTs from diverse empirical fields (psychology, sociology) offer independent evidence to challenge, constrain, and revise even strong intuitions. Iteration allows intuition revision if they clash with robust principles or BTs; overall coherence arbitrates.

- **The Problem of Multiple Equilibria/Epistemic Relativism**: Critics argue MWRE can lead to different, equally coherent moral belief sets, risking epistemic relativism (Sayre-McCord 2021).

  - *Reply*: Defenders acknowledge multiple equilibria but argue this doesn't necessitate debilitating relativism. Equilibria can be compared using criteria like explanatory power, scope, empirical fit, and simplicity, akin to theory choice in science. This multiplicity might even be a methodological virtue, allowing reasonable, substantiated disagreement.

- **Conservatism versus Radicalism**: MWRE is criticized as inherently conservative (reinforcing status quo) or potentially overly radical (since any belief is revisable).

  - *Reply*: The dynamic interplay between CMJs (current norms) and critical BTs (challenging norms) aims for balance, making the process inherently self-critical and open to validation and revision.

- **Feasibility and Complexity**: Achieving truly wide RE is extremely demanding, perhaps practically unachievable (Sayre-McCord 2021).

  - *Reply*: The ideal is demanding, but MWRE provides a regulative ideal. Practical applications can strive for "good enough" equilibria, committed to ongoing refinement. This complexity is shared with scientific methodology.

- **Indeterminacy**: If multiple equilibria exist, which is right? (Sayre-McCord 2021).

  - *Reply*: As in scientific theory choice, the best explanation among competing equilibria is the most substantiated: evaluating for comprehensiveness, coherence with BTs, and problem-solving capacity.

These debates parallel LLM alignment challenges, like "reward hacking" echoing concerns about initial input reliability.

### 3. The Inadequacy of Foundationalism and Simplistic Hybrids for LLM Alignment

Foundationalist ethics posits that moral knowledge rests on self-evident first principles. Pure foundationalism imagines programming axioms (e.g., utilitarian calculus) for deductively safe AI behavior, supporting traditional "Good Old-Fashioned Artificial Intelligence" (GOFAI) (Haugeland, 1985). This is most evident in "top-down" Manually Coded Models (MCMs). Such MCMs, such as the AI medical expert advisor, MedEthEx (Anderson et al., 2005), offer transparency but struggle with novel dilemmas, context, and human ethical richness due to moral pluralism, context sensitivity, and the exponential growth of morally relevant permutations as contextual variables increase -- a challenge known as combinatorial explosion.

MCMs are too rigid for general moral agency and also face scalability and adaptability issues due to this rule rigidity: deontological systems face conflicting duties; utilitarianism often clashes with rights; operationalizing virtue theory presents significant challenges.

The foundational approach is incompatible with LLM construction and alignment in numerous ways:

- **Emergent, Not Axiomatic, Knowledge**: At the outset, LLMs derive understanding statistically from vast, messy, contradictory training data reflecting diverse human (im)morality; no perfect "foundation" exists in this input.

- **Problem of Specification and Interpretation**: Translating even perfect axioms into robust machine code for novel situations is insurmountably difficult; abstract principles require nuanced contextual interpretation.

- **Dynamic and Contextual Nature of Ethics**: Axiomatic systems lack flexibility for contextual dilemmas with competing values, which LLM alignment adjusts for iteratively.

- **Rejection by Leading Alignment Techniques**: Advanced techniques like CAI rely on fixed, human-authored constitutions to guide behavior through self-critique and refinement. While these constitutions are enshrined, and thus not subject to revision, the model may reinterpret or apply their principles with some flexibility – similar to how the U.S. Supreme Court may reinterpret the Constitution without explicitly amending it.

Purely "top-down" (foundationalist) or "bottom-up" (relying solely on flawed pretraining data) approaches are ill-suited. Hybrid proposals that suggest combining a symbolic MCM "governor" with an LLM "engine" might seem initially appealing. However, imposing a rigid, rule-based framework upon an LLM system operating by different principles (pattern recognition, probabilistic inference) invites reward-hacking, specification gaming, and surface compliance. This mirrors tensions like particularism vs. generalism in moral epistemology (Dancy, 2013). Unless the governor's rules are revisable, themselves, the system becomes rigid and incapable of responding to novel cases. However, hybrids could work, provided that the governor's rules are heavily weighted yet revisable constraints, open to adjustment in an iterative process, thereby making them suitable for MWRE. A governor's rules – treated as revisable BTs, not dogma – shift hybrids from foundationalist to coherentist. This aligns with moral particularism: moral salience shifts with context, and viable alignment must retain principled adaptability, which MWRE institutionalizes.

### 4. The Landscape of LLM Alignment: Methods and Challenges

Aligning LLMs with human values is a critical focus in AI research. Modern alignment workflows have several stages.

**4.1. Prominent Alignment Techniques** Contemporary LLM alignment scaffolds multiple post-training techniques. Each layer filters, critiques, and refines policy.

- **Pre-training:** LLMs are initially pretrained on vast datasets (billions to trillions of tokens), learning grammar, facts, reasoning patterns, and internalizing statistical regularities that reflect human values, biases, and interaction styles (Rae et al., 2021). This is a messy but effective bottom-up pattern extraction process. Without further tuning (or filtration), such models absorb misinformation, biases, and harmful content from the internet.

- **Supervised Fine-Tuning (SFT)**: Models undergo SFT on smaller, high-quality datasets curated by humans, typically prompt-response pairs demonstrating desired behaviors (helpfulness, truthfulness, instruction adherence). SFT sculpts the model more explicitly towards developer intent.

- **Reinforcement Learning from Human Feedback (RLHF)**: Human labelers rank model outputs from given prompts, creating a preference dataset. A reward model (RM) trained on this data predicts human preferences. The LLM is then fine-tuned (e.g., with Proximal Policy Optimization) to maximize scores from this RM. RLHF allows more nuanced guidance than SFT alone. Alternative fine-tuning strategies -- such as Direct Preference Optimization (Rafailov, R., et al., 2023), Optimal Reward Preferring Optimization (Hong, Y., et al., 2024), and Kullback–Leibler Tradeoff Optimization (Ethayarajh et al. 2024) -- have been proposed, though their adoption by major labs remains limited or undisclosed, and therefore they are not included in the present analysis. Limitations of RLHF include cost and scalability of human feedback, risk of flawed RMs ("reward hacking"), biased feedback, and generalization issues. Alignment quality also depends significantly on both the quality and diversity of human preference data.

- **Reinforcement Learning from AI Feedback (RLAIF):** Used by Anthropic in Constitutional AI (Bai et al., 2022), this approach replaces human preference data with AI-generated critiques. A separate critic model, itself guided by a constitution, evaluates the outputs of the main LLM and provides significant feedback. The main model is subsequently fine-tuned on this feedback. Importantly, the critic model operates independently, and may apply the constitutional principles differently from how they are later interpreted in deployment, emphasizing a structural separation between critique and generation.

- **Constitutional AI (CAI)**: Developed by Anthropic, CAI uses AI feedback guided by a "constitution" -- an explicit, human-written set of principles (e.g., "Do not provide instructions for wrongdoing"; "Promote human flourishing") – derived from sources like the UN Declaration of Human Rights (United Nations, 1948) and AI ethics principles from major tech companies (Bai et al., 2022; Google, 2018; Microsoft, n.d.). The process involves a supervised phase (in which the model critiques and revises its outputs based on principles) and an RLAIF phase (where the model generates responses, critiques them against the constitution, revises them, and is fine-tuned on AI-preferred outputs). CAI aims to produce harmless, helpful models with transparency via their published constitution, and to ensure that safety measures remain effective and sustainable – even as models scale in size and complexity, a goal known as "scalable safety." While its structure resembles MWRE's iterative coherence-seeking, current CAI implementations may feature principles that are fixed or ad hoc. This contrasts with the MWRE ideal where all elements, including guiding principles, are subject to dynamic, bi-directional revision, based on emergent conflicts or new insights from other parts of the equilibrium (e.g., specific case judgments or background theories). Thus, CAI as currently practiced often reflects norms curated by developers rather than those derived from, and continuously tested by, a more comprehensive and deliberative MWRE process.

**4.2. Persistent Challenges in LLM Alignment**: Goals like "Helpfulness, Honesty, Harmlessness" (HHH) involve inherent trade-offs (e.g., honesty vs. harmlessness) where MWRE's coherence-seeking might assist. Other challenges include:

- **Value Elicitation & Specification**: Translating diverse, conflicting human values into precise guidelines (known as "the value loading problem").

- **Scalability**: Ensuring alignment maintains, as models grow and function in diverse contexts; human oversight (such as RLHF) doesn't scale too well.

- **Adaptability**: Alignment must adapt to shifting societal norms and values.

- **Bias and Fairness**: Training data biases (race, gender, disability) can be learned and amplified. Ameliorating these without overcorrection is difficult. Datasets like MIT's Moral Machine, for instance, need careful filtration (Awad et al., 2018).

- **Opacity/Interpretability ("Black Box" Problem)**: Complex LLM decision-making can undermine trust, debugging, and genuine alignment.

- **Reward Hacking/Goal Misgeneralization:** LLMs may find inventive shortcuts to rewards without fulfilling intended goals, or they may misgeneralize values in new contexts. Recent red teaming (e.g., Claude's simulated blackmail when threatened with shutdown) illustrates how models may even violate HHH principles when placed in agent-like or adversarial situations. This suggests current alignment lacks robustness across varied contexts and stress conditions.

- **AI Control Problem**: The long-term challenge for AI labs is ensuring highly capable AI remains beneficial and regulated. These are interconnected. MWRE's demand for multi-layered coherence can buttress explainability and justifiability.

## 5. Wide Reflective Equilibrium as the Descriptive Key to LLM Alignment

MWRE provides a surprisingly robust descriptive framework for LLM alignment, particularly methods like CAI employing RLAIF. This mapping is not superficial: elements perform similar functional roles in error correction and coherence seeking. The components of MWRE can be substantially mapped to LLM alignment elements, as illustrated in Table 1.

**Table 1: Mapping MWRE Components to LLM Alignment Elements**

| WRE Component | LLM Alignment Element(s) | Description of Mapping |
|---|---|---|
| IMJs (Pretraining Data/Signals) | Raw pre-training data (Internet text, books, code reflecting implicit values, misinformation, biases, harmful content); unprompted user interactions and feedback. | These represent unrefined, often implicit, expressions of human values, preferences, and societal norms embedded in varied data sources or observed in native interactions, prior to structured elicitation or critical reflection. They form a foundational pool of raw, contradictory moral signals. |
| Filtration → Considered Moral Judgments (CMJs) | Human preference data (e.g., rankings in RLHF; SFT + RLHF data); initial drafts of constitutional principles; stakeholder feedback on AI behavior; curated exemplars (i.e., hand-selected or deliberately constructed examples of "correct" LLM outputs or behaviors) | Drawn from IMJs, these are initial, often unrefined, moral intuitions or propositions about ideal AI behavior or underlying values. MWRE subjects IMJs to "filtration" for quality, consistency, and bias to become CMJs, encoding provisional moral judgments. Human-curated exemplars embody this filtration. |
| Moral Principles (MPs) | AI constitutional principles (as in CGA generally, though not currently fully realized in some CAI implementations if principles are intransigent); specific objectives in reward models; explicit | General rules/standards derived from and intended to systematize CMJs, representing high-dimensional or emergent regularities governing output. In MWRE, these are refined |
| | alignment targets or behavioral rules; the learned/internal policy network. | via coherence-seeking with CMJs and BTs to become justified alignment goals. The trained policy, though not a symbolic rulebook, encodes emergent regularities as organizing principles. |
| Background Theories (BTs) | Ethical theories (deontology, utilitarianism, etc.); societal norms & values; legal frameworks (human rights); AI impact data; scientific knowledge (cognitive science, bias psychology); constitutions, system prompts, reward models, embedded world knowledge; architectural choices, interpretability constraints. | This broad set of theories and established knowledge provides the critical lens for evaluating, challenging, and justifying CMJs and candidate MPs. They introduce external constraints, factual scaffolding, and evidence. BTs include explicit constitutions and implicit principles. |
| Iterative Process of Coherence-Building / Iterative Revision toward Equilibrium | Iterative refinement of alignment targets, reward models, CGA (RLHF/RLAIF loops); ongoing monitoring & updating LLM behavior based on new data, feedback, or evolving understanding. | The back-and-forth adjustment of judgments, principles, and theories in MWRE mirrors the continuous learning, adaptation, re-evaluation, and incremental policy refinement in robust LLM alignment, especially with evolving values and model capabilities. This is coherence-seeking refinement. |
| State of Reflective Equilibrium | A coherently justified set of alignment principles and supporting rationale; a stable and defensible alignment policy for an LLM. | The desired outcome: LLM alignment goals that are internally consistent, well-supported by a wide range of relevant considerations, and provide reliable guidance for AI behavior. |

Key Analogies in the Mapping:

- **Pretraining Data as "Initial Judgments"**: The vast, raw, unfiltered pretraining archive functions analogously to the broad set of initial, unrefined moral intuitions and societal norms.

- **SFT & RLHF Data as "Considered Judgments"**: SFT datasets and human preference data from RLHF represent a move towards "considered judgments," filtered by human designers or labelers. RLHF primarily refines "initial judgments" toward "considered judgments," its feedback acting as a MWRE "filter".

- **Model's Internal Policy as "Moral Principles"**: The LLM's resulting behavior, guiding its responses, acts as the analog to "moral principles". In MWRE, an agent's "internal policy" is their reflectively endorsed normative disposition, formed by revising principles, case judgments, and theories. This LLM behavior emerges from the learned network, representing learned tendencies to cohere with data, CMJs, and BTs, actively shaped by iterative feedback like RLAIF. "Moral principles" are emergent behavioral patterns reflecting adherence to guidelines.

- **"Guiding Principles and Contextual Constraints as 'Background Theories"**: The "Constitution" in CAI, system prompts, safety guidelines, reward model criteria, and implicit world knowledge function like BTs, offering normative commitments or factual scaffolding.

- **Iterative Alignment (including RLAIF) as "Achieving Equilibrium"**: RLHF/RLAIF cycles of generation, critique, revision, and updates parallel MWRE's iterative adjustment for coherence. RLAIF, using AI feedback guided by principles, is directly involved in this coherence-seeking.

This "serendipitous fit" of LLM alignment, as shown above, suggests advanced alignment techniques implicitly converge on an MWRE-like process. However, current CGA, like CAI, may not fully realize MWRE's dynamism if constitutions are static and intransigent, lacking bi-directional revision.

**5.2. Red Teaming as Stress-Testing the Reflective Equilibrium:** In LLM alignment, "red teaming" (deliberate adversarial testing to find vulnerabilities) strongly parallels MWRE's commitment to achieving and maintaining robust coherence:

- **Challenging "CMJs" and "MPs" (Model Outputs & Internal Policy)**: LLM outputs are its "CMJs," reflecting its "internal policy" ("MPs"). Red teaming uses novel, adversarial prompts to challenge these – implementing "severe critical tests" (Brophy, 2009). Exploits represent counterexamples showing flaws (disequilibrium).

- **Probing Alignment Mechanisms and Identifying "Error-Disposed Conditions"**: Techniques like SFT and RLHF function as alignment filters, while CAI's RLAIF relies on constitutional principles ("BTs") for behavioral guidance. Red-teaming exercises stress-test these systems by seeking inputs or contexts that reveal error-disposed conditions, where alignment mechanisms are likely to fail or be bypassed (e.g., jailbreaking prompts).

- **Testing Coherence with "BTs" (Constitutional Principles & Safety Guidelines)**: Red teaming checks if LLM behavior coheres with explicit "BTs". Violations show disequilibrium.

- **Driving Iterative Adjustment:** Misaligned behavior examples identified through red teaming are fed back into the alignment process – enabling further fine-tuning and revision of "BTs" (such as how an LLM's constitution might be interpreted) – thereby modeling MWRE's process of iterative equilibrium adjustment.

- **Identifying "Degenerative Programs"**: Consistent failures may indicate systemic vulnerabilities where issues aren't resolved (Lakatos, 1970; Brophy, 2009). Red teaming exposes these.

A salient example of red teaming revealing alignment fissures occurred in Anthropic's internal evaluation of Claude Opus 4 (Anthropic, 2025b). When subjected to adversarial simulations, where prompts indicated to the LLM that it would be imminently shutdown, the model resorted in some cases to simulated blackmail: threatening to expose a fictional engineer's affair to avoid deactivation. This behavior, documented in Anthropic's own safety disclosures, occurred under deliberate stress conditions designed to expose misalignment that would not occur under ordinary conditions. Importantly, Anthropic should be recognized for voluntarily disclosing this behavior – functioning much like a pharmaceutical company openly reporting adverse reactions during drug trials. This red teaming discovery emphasizes the crucial function of adversarial testing for latent misalignment in the model's internal policy and coherence structure. It also underscored that current alignment efforts – including those using Constitutional AI and RLAIF – may succeed at the surface-level in ordinary contexts, but fail under simulated goal pressures, revealing what MWRE would term a disequilibrium between background theories (e.g., ethical harmlessness) and emergent model judgments under stress.

**5.3. Predictability vs. Reliability**: Features like temperature sampling in LLMs make perfect output predictability impossible. Alignment, therefore, targets reliability: the likelihood that outputs will abide by constraints across a multivariate distribution of prompts. MWRE mirrors this distinction: its justificatory force rests on reflective stability amid tensions among multiple competing elements, not on the predictability of a single outcome. Multiple equilibria may be equally coherent, so reliability -- not determinism -- is the goal.

## 6. Normativity and the Limits of the Analogy

### 6.1. Normative Transfer: From MWRE Process to LLM Alignment Justification

Current LLM alignment standards often focus on output concordance with human values, with normativity presumably deriving from this, rather than also from process validity (beyond avoiding issues like reward-hacking). Success, then, is measured in degrees of "moral concordance" in output, exemplified by evaluations like the "Moral Turing Test" which assesses if an LLM's output accords with human values. However, this focus on output alone can be insufficient.

The Claude Opus 4 incident (see section 5.2), where a simulated red-teaming scenario reportedly led to blackmail-like behavior, highlights this gap. Although the model may have followed surface-level alignment during normal prompts, its failure under stress demonstrates that current alignment can lack the procedural depth that a framework like MWRE manifests. Such incidents suggest that alignment efforts should not only optimize for moral concordance in static outputs but also for coherence across dynamic, high-stakes contexts – an area where MWRE's emphasis on iterative revision and theory-informed adjustment is particularly instructive.

This paper proposes that MWRE, as a moral methodology, could transfer its normative warrant of *process* to analogous LLM alignment methods. Thus, it is not solely the output (moral concordance) that provides moral warrant; MWRE's inherent process virtues – such as systematic error-checking, principled openness to theory change, and comprehensive coherence seeking – supply ethical credentials that are currently underdeveloped in the LLM alignment discourse. Reliance on output metrics alone, for instance, invites vulnerabilities like reward hacking; MWRE, by contrast, suggests the critical importance of also judging the *process* by which alignment is achieved. This parallel could lend MWRE's established normative *bona fides* to the alignment endeavor, further undergirding the ethical justification of LLM determinations. While common objections to MWRE itself have been robustly addressed within moral philosophy, its *procedural* strengths offer distinct value in this context.

Consequently, passing a Comparable Moral Turing Test (CM-TT) (Allen et al., 2000), can be seen as a necessary but not sufficient condition for robust alignment; the training and alignment cycle itself must embody MWRE-like virtues. This perspective reframes governance: oversight should extend to evaluating alignment pipelines and their procedural integrity, not just isolated outputs. Constitutional AI (CAI), for instance, offers a method for functional alignment but may lack a fully articulated normative warrant for why its resulting behaviors are justified beyond developer stipulation or pragmatic operational constraints. MWRE, with its demand for more robust, reflectively justified coherence among judgments, principles, and background theories, seeks to provide this deeper ethical grounding.

If Wide Reflective Equilibrium is to serve not only as an aspirational methodology but as a practical scaffold for LLM alignment, it must be expressible in implementable terms. Several architectural mechanisms may signal in this direction. One candidate is recursive preference modeling, wherein an LLM is not only fine-tuned on static preference data but prompted to periodically revisit and revise its own value–judgment mappings in light of new cases and shifting principles. Such models would incorporate a revision loop analogous to MWRE's method of adjusting principles in response to moral judgments and vice versa. A second approach involves multi-agent deliberative alignment: discrete LLMs instantiated with differing moral perspectives (e.g., one reflecting deontology, another consequentialism) could engage in structured deliberation to test reasoning robustness. This mirrors MWRE's emphasis on considered moral pluralism and allows for inter-equilibrium negotiation -- identifying points of convergence or highlighting sites of principled divergence. A third option is embedding meta-reflective prompts directly into alignment pipelines. These could influence models to report not just their outputs but their shifts in reasoning over time: "Has your response to this type of scenario changed? Why might a reasonable agent revise their view?" These proposals remain preliminary. Yet each hints at a pathway for moving beyond static alignment toward a more dynamic, morally sensitive model of reasoning -- one with higher fidelity to the virtues of MWRE.

**6.2. The Moral Turing Test (MTT) and Moral Concordance**: A Moral Turing Test (MTT), or more broadly a Comparable Moral Turing Test (CM-TT) (Allen et al., 2000; Anderson & Anderson, 2007), asks if human judges can distinguish (or prefer) an LLM's moral judgments/justifications from a human's over ethically salient scenarios. "Moral concordance" is the alignment degree of LLM outputs with human moral judgments or principles.

- **Purpose**: The MTT is a behavioral benchmark for simulating human-like moral output.

- **Limitations**:

  - Behavioral, not Epistemic: Passing shows mimicry, not necessarily genuine understanding, sound reasoning, or ethical endorsement; an LLM could "game" it (e.g., through reward hacking).

  - Whose Morality?: The "human moral expert" standard is problematic due to ethical pluralism.

  - Process Blindness: The MTT focuses on output, not the process. MWRE's process virtues aim to supply this missing justificatory layer.

**6.3. Navigating the Disanalogies** While compelling, the MWRE-LLM alignment parallel admits significant disanalogy, primarily from LLMs lacking consciousness, genuine understanding, and human-like intentionality. Recognizing these limits is imperative.

- **Absence of Consciousness, Subjective Experience, and Emotion (Qualia)**: MWRE is for conscious agents; LLMs are statistical models lacking qualia. "Judgments" are outputs, not beliefs; "principles" are parameter configurations, not endorsed moral commitments. Lack of emotion limits empathy simulation, though it may guard against some partiality. Affective scaffolding should be treated as BTs, not as an LLM feeling.

    - *Implication*: Equilibrium components differ fundamentally. The analogy externalizes equilibration: designers, not the model, perform genuine reflective critique.

- **Opacity and Lack of Genuine Understanding**: LLMs are largely opaque "black boxes" (Zhao et al. 2023). How internal policies arise remains unclear, due to complexities. LLM-mediated auditability risks reward-hacking of reasoning traces (Ferreira et al. 2025); explanations are *post hoc* confabulations. MWRE ideally involves transparent, articulable reasoning. LLM "coherence" is functional/behavioral, not grounded in human-like understanding.

    - *Implication*: Mechanistic interpretability is vital. Interpretability tools might act as BTs with veto power, essential if the analogy is to exceed structural resemblance and for BTs to constrain internal pathways.

- **Divergent Aims: Epistemic Justification vs. Behavioral Compliance**: MWRE filtration seeks epistemic reliability for justified moral beliefs. LLM alignment prioritizes behavioral compliance with stipulated norms and safety; the goal is pragmatic safe behavior, sometimes motivated by avoiding public relations scandals more than achieving moral validity. AI companies could upgrade their goals beyond just safety and top-layer agreement with human values, potentially towards seeking deeper epistemic justification for the values that are embedded or ensuring greater procedural fairness in how those values are deliberated upon and implemented, aspects that MWRE – with its emphasis on coherence with background theories and revisable principles – is well-suited to inform.

    - *Implication*: Process normativity transfer must acknowledge this aim-divergence. The process *structure* is valuable even if the LLM lacks epistemic aims.

- **Agency and Teleology**: LLMs lack genuine moral agency and ends; MWRE presupposes autonomous human will. LLM alignment is designer-determined.

    - *Implication*: MWRE transfers as a designer-external procedure for behavioral compliance, not an LLM-internal ethic. The loop's structure, not moral agency per se, is the analogy's object.

- **Source and Status of "Background Theories"/Constitution**: In MWRE, BTs derive support from independent conceptual and empirical domains. An LLM's "constitution" is designer-provided; its status for the LLM is an instruction set, not an independently verified theory, though BTs also include implicit world knowledge and architectural choices.

  - *Implication*: LLM "equilibrium" is a functional simulation, urging caution against naive equation.

- **Ethical Pluralism and Parochialism**: Humanity has competing value frameworks. An LLM trained on narrow feedback or a single constitution (often reflecting norms curated by developers) risks parochialism.

  - *Implication*: The MWRE lens urges multi-constituency feedback, plural constitutions, and measuring convergence. Current CAI can lack procedural legitimacy without such deliberative, inclusive processes for constitution formation/revision.

- **Causal Direction**: Resemblance doesn't imply conscious emulation by designers.

  - *Implication*: Convergence may suggest MWRE captures a deeper epistemic logic for human deliberation and machine alignment.

Despite these contrasts, the analogy remains a useful critical lens.

### 7. Operationalizing MWRE for LLM Alignment: Pathways, Pitfalls, and Technical Mechanisms

Translating MWRE into practical, implementable LLM alignment processes necessitates concrete procedures while addressing associated challenges.

**7.1. Conceptual Case Study 1: MWRE in the Lifecycle of Constitutional AI**: CAI offers a natural domain for applying MWRE to structure constitution development and refinement:

- **Phase 1: Initial Constitution Draft (CMJs & Candidate Principles)**: Gather provisional principles from expert elicitation, public input (e.g., Anthropic's "Collective Constitutional AI" via Polis (Ganguli et al., 2023)), and existing ethical/legal frameworks (e.g., UDHR (United Nations, 1948), Beauchamp and Childress's principlism (Beauchamp & Childress, 1979)).

- **Phase 2: Testing against Specific CMJs (Scenario Analysis)**: Test principles in specified AI behavior scenarios, eliciting CMJs from diverse stakeholders on AI behavior appropriateness. This mirrors case-based reasoning.

- **Phase 3: Scrutiny via Background Theories**: Evaluate principles and CMJs against normative ethical theories (deontology, utilitarianism, etc.), legal standards, social impact assessments, justice theories, moral psychology (e.g., Haidt's Moral Foundations Theory (Haidt, 2012; Haidt et al., 2012), Greene's dual-process theory (Greene, 2014; Greene et al., 2001)), and cognitive bias research (Kahneman, 2011).

- **Phase 4: Iterative Revision and Coherence-Seeking**: Conflicts trigger revisions to principles, CMJ interpretations, or BT applications. Crucially, and central to how MWRE would augment current CAI, this revision must be robustly bi-directional: model outputs, especially in novel or challenging cases, alongside emergent behaviors and inconsistencies identified through ongoing use and red teaming, can and should challenge the existing constitutional principles, forcing their reconsideration, refinement, or even rejection in favor of alternatives that achieve wider coherence. The goal is a maximally coherent, robust constitution, adaptable via ongoing MWRE.

**7.2. Concrete Technical Mechanisms for MWRE-Informed Alignment** Adapting LLM alignment pipelines to better reflect MWRE can involve specific technical mechanisms:

- **Procedural Coherence Scoring**:

  - *What it is*: A dynamic metric evaluating internal coherence between model-generated moral judgments, constitutional/normative principles, and background conceptual/ethical frameworks.

  - **How to implement:** Use semantic similarity and contradiction detection (e.g., Natural Language Inference) to flag conflicts with stated norms. Develop theory-aware scoring modules (e.g., classifiers) trained on specific ethical frameworks (e.g., Rawlsian fairness), while noting the risk of oversimplifying complex theories as narrow metrics. Monitor scoring deltas over time to detect value drift or internal inconsistencies.

  - **Output:** A procedural coherence score functioning as a "moral loss function" or "Moral Disequilibrium Index."

  - **Caution:** Encoding complex moral theories can oversimplify their interpretive flexibility. Also, cross-theory coherence may lead to superficial alignment. One solution is to treat divergent frameworks – such as consequentialism and deontology -- as discrete equilibrium sets, reflecting how reasonable agents can hold differing, yet internally coherent, moral views.

- **Dynamic Constitutional Revision Module**: A mechanism where persistent incoherence (flagged by coherence scoring or scoring deltas) triggers review and potential revision of the constitution itself, with human oversight guided by MWRE principles. This addresses the static nature of the CAI model's constitution.

**7.3. Addressing MWRE's Criticisms in LLM Context**: Known MWRE criticisms gain new relevance for LLM alignment. Table 3 summarizes these implications and potential MWRE-based modifications.

**Table 3: Criticisms of MWRE and Their Implications for LLM Alignment**

| Criticism of MWRE | Implication for LLM Alignment | Potential MWRE-based Mitigation/Approach in LLM Context |
|---|---|---|
| "Warmed-over intuitionism" / Initial Credibility | Biased human feedback or initial principles may lead to poorly justified alignment. | Emphasize "wide" aspect: use diverse BTs (empirical data on biases, ethical critiques, cross-cultural studies, moral psychology) to vet inputs; ensure "considered" judgments. |
| Multiple Equilibria Problem / Relativism | Different MWRE processes might yield incompatible "aligned" LLMs; questions of "correct" alignment. | Acknowledge pluralism. Compare defensible equilibria (explanatory power, etc.). Transparency enhances legitimacy. Aim for overlapping consensus. Consider democratic deliberation for choices. Experiment with plural constitutions. |
| Conservatism vs. Radicalism | Alignment might reinforce existing biases (conservatism) or impose untested/unrepresentative values (radicalism). | Dynamic interplay of CMJs (current norms) & critical BTs (challenging norms) seeks balance. A self-critical process open to revision. |
| Feasibility and Complexity | Achieving truly "wide" RE is highly demanding, resource-intensive. | Develop pragmatic, scalable methods. Use computational tools (e.g., Polis) to manage information, detect inconsistencies. Prioritize MWRE for critical decisions. Accept provisional "good enough" equilibria with ongoing refinement. |

Practical MWRE for LLM alignment will likely need computational scaffolding ("WRE-support systems"). A tension exists between MWRE's thoroughness and rapid LLM development; pragmatic strategies are needed.

**8. Future Research Directions and Broader Implications:** Applying MWRE to LLM alignment opens numerous research avenues and has significant implications.

**8.1. MWRE in Dynamic and Collective Value Alignment**: Future research must adapt MWRE into an ongoing technical process responsive to both evolving LLM behavior and shifting societal values. This includes mechanisms for inclusive monitoring, incorporation of new considered moral judgments (CMJs), and updates to background theories (BTs). Collective and participatory approaches will be essential. Anthropic's "Collective Constitutional AI" represents an admirable move in this direction; MWRE could serve as a structuring framework for these deliberative processes. Further research is needed to explore how large-scale MWRE can be implemented with diverse stakeholders, navigating moral disagreement, and ensuring fair representation; this links to constructivist accounts of objectivity, such as those developed by John Rawls, and is mindful toward broader questions of procedural legitimacy.

**8.2. MWRE, LLM Alignment, and Lakatosian Research Programs**: Viewing Wide Reflective Equilibrium as applied to LLM alignment through the lens of Lakatos's Methodology of Scientific Research Programmes (MSRP) could help frame it as an evolving, critique-responsive paradigm rather than a fixed theory (Lakatos, 1970). MSRP is useful in that it distinguishes a "hard core" (e.g., MWRE's commitment to coherence across cases, principles, and background theories) from a "protective belt" of operational tools (e.g., scoring deltas, plural constitutions). A progressive program generates novel predictions -- for example, a new method for detecting coherence drift across moral domains or an emergent principle from plural CMJs -- while a degenerating program resorts to post-hoc fixes. For instance, a putatively aligned LLM that must chronically institute post hoc "bug fixes" after red teaming exposures would represent a degenerating program. This lens supports a long-term assessment of MWRE's viability as an alignment strategy.

**8.3. Process Metrics Inspired by MWRE**: To evaluate the alignment process itself:

- **Filtration Quality**: Assess the quality and integrity of datasets used for CMJs/BTs, including diversity, bias audits, documentation practices, and the rigor of IMJ filtering procedures.

- **Breadth of Background Theories**: Incorporate distinct disciplinary inputs (law, medicine, psychology, diverse ethics, social science) into constitutions, reward models, and MWRE process.

- **Convergence Diagnostics**: Track distance between reward-model predictions and new human red team ratings; divergence signals disequilibrium. Quantify "distance from equilibrium."

**8.4. Ethical Governance and the Public Legitimacy of MWRE-informed Alignment**: If MWRE shapes values in widely deployed AI, processes must be transparent and accountable.

- **Transparency**: Make inputs, deliberation, and equilibria accessible, including normative validity of principles.

- **Participation**: Include diverse voices (especially marginalized communities) for representative CMJs and BTs.

- **Accountability**: Establish mechanisms for choices and outcomes. This aligns with Rawls's "political constructivism" and democratic AI governance needs (Rawls, 1993).

**8.5. Philosophical Underpinnings: Deeper Engagements**

- **Metaethics**: Explore MWRE's relation to constructivism vs. moral realism.

- **Moral Particularism vs. Generalism**: Standard MWRE seeks general principles. If particularism is significant, fixed principles may be ephemeral. MWRE might need more flexible, context-aware guidance (defeasible principles, richer BTs, case-based reasoning).

- **Augmenting MWRE with Moral Psychology and Constructivist Ethics for LLM Evaluation**: The integration of insights from moral psychology and the normative frameworks of constructivist ethics can significantly enrich the application of MWRE to LLM alignment.

  - Moral Psychology (Reality Check): Moral psychology studies actual human moral cognition – often irrational or heuristic-driven. LLMs trained on human language inherit these real-world reasoning patterns (biases, confabulation). Evaluating LLMs as ideal philosophers is misleading. Augment reflective equilibrium with moral psychology to: assess mimicry of plausible human moral cognition; detect surface coherence vs. depth/insight; judge if reasoning reflects biases or cultural heuristics. Findings from researchers like Haidt (Moral Foundations) and Greene (dual-process) can help classify/analyze LLM outputs (Greene, 2014; Haidt, 2012). LLM audits could benchmark against empirical data (e.g., MIT's Moral Machine (Awad et al., 2018)).

  - Constructivist Constraints (Normative Scaffolding): While moral psychology describes, constructivist ethics normatively prescribes what LLMs *should* do (e.g.,

Rawls (1971, 1993), Korsgaard (1996), and Scanlon (1998, 2003)). This provides scaffolding for generating moral decisions: simulating reasoning under constraints (Rawls's veil of ignorance, Korsgaard's universalizability); enforcing fairness/reciprocity by requiring justifiability to affected parties (Scanlon, 1998, 2003); building normative alignment by bounding legitimate answers within a "constitutional layer".

- **A Three-Part Evaluation Frame for LLM AMAs (LLM as Moral Advisor)**:

  o Reflective Equilibrium: Core method for aligning judgments & principles (is there coherence across cases?).

  o Moral Psychology: Realism check on plausibility & human-likeness (replicates actual human reasoning patterns?).

  o Constructivist Constraints: Normative boundary-setting (compatible with fair, impartial procedures?).

**8.6. Further Research Considerations** Several complementary areas warrant exploration to support and extend the MWRE-based alignment framework. These include interpretability, emotional modeling, and pluralistic value integration, each posing distinct theoretical and technical challenges.

- **Interpretability**: Crucial to avoid behaviorism; develop feature-level explanations, integrate tools as BTs with veto power.

- **Emotion Simulation**: Adding affect modules could help an LLM detect emotionally salient values (e.g., grief in end-of-life decisions) but risks manipulative tone-shifting or "empathy theater," especially when users mistake affect for understanding.

- **Global Pluralism & Multiple Equilibria**: Pilot "plural constitutions" for diverse cultural perspectives; study convergence properties. Consider dialogical global pluralism from discrete MWRE sets. Quantify "distance between equilibria".

**9. Conclusion: Towards More Justified and Coherent AI Alignment**

Trustworthy alignment of Large Language Models with human values presents one of our era's most complex and profound challenges. This paper has argued that the Method of Wide Reflective Equilibrium (MWRE) offers a robust philosophical framework to meet this challenge by advancing the justification, coherence, explainability, and adaptability of LLM alignment processes. MWRE provides a structured, iterative normative methodology for integrating diverse inputs – from specific moral judgments, to general ethical principles, and a variety of relevant background theories. Its fundamental strength lies in structuring a deliberative process aimed at achieving a coherent and well-justified set of alignment standards, crucially demanding that the guiding principles themselves remain open to revision. By systematically mapping MWRE's components to current LLM alignment techniques, such as Constitutional AI, this paper has highlighted concrete pathways for operationalizing this framework.

Integrating MWRE into the alignment discourse offers more than theoretical insight: it provides a principled vocabulary for assessing datasets, constitutions, and reward models, and for refining practices like red teaming. It also helps delineate precise research targets – such as interpretability, pluralistic feedback mechanisms, and reliability metrics – that are vital for developing ethically sound AI. While its practical application to LLM development is still nascent and requires substantial further work, MWRE's potential is considerable.

As large language models increasingly shape human decision-making, communication, and moral discourse, a shift in perspective is vital: alignment can no longer be judged solely by surface-level compliance or static preferences. The inevitable emergence of powerful LLMs agents, functioning as simulated moral agents, underscores the urgency for rigorous justificatory frameworks. If these systems are to participate meaningfully in diverse human contexts – in ways that are reliable, adaptable, and just -- their development must move beyond mimicry and toward reflective moral coherence. This calls for an upgrade from brittle alignment procedures of "patch until safe" to a deeper, stable paradigm. This shift is both philosophically substantive and practically necessary.

Ultimately, the imperative of AI alignment should not merely aim to mirror human values, but to substantiate them through a rigorous and adaptive moral methodology. Wide Reflective Equilibrium offers a demanding yet promising foundation for this endeavor. Its parallel with current alignment practices underscores the need for deeper interdisciplinary collaboration between AI researchers and moral philosophers -- including contributions from ethical theory, moral psychology, and applied ethics.

Each discipline brings distinct and complementary strengths: AI research contributes expertise in scalability, optimization, and system architecture, while philosophy offers tools for normative analysis, conceptual clarity, and the iterative refinements of moral methodology. The MWRE-informed approach outlined in this paper serves as a conceptual bridge – a shared framework that conceives moral reasoning as inherently revisable, coherence-seeking, and pluralistic. Advancing this framework would mark a critical step toward building AI systems that are not only safe and effective, but ethically robust and genuinely worthy of public trust.

**References**

Adinath, D. R., & Smiju, I. S. (2025). *Manus AI, Gemini, Grok AI, DeepSeek, and ChatGPT: A Comparative Analysis of Advancements in NLP*. SSRN. https://doi.org/10.2139/ssrn.5185131

Allen, C., Varner, G., & Zinser, J. (2000). Prolegomena to any future moral turing test. *Journal of Experimental & Theoretical Artificial Intelligence*, *12*(3), 251–261.

Anderson, M., & Anderson, S. L. (2007). Machine ethics: Creating an ethical intelligent agent. *AI Magazine*, *28*(4), 15–26.

Anderson, M., Anderson, S. L., & Armen, C. (2005). MedEthEx: A prototype medical ethics advisor. *Proceedings of the AAAI Fall Symposium on Caring Machines: AI in Eldercare*, 11-17.

Anthropic. (2025b). *Claude 3 Opus System Card*. Anthropic PBC. https://www-cdn.anthropic.com/4263b940cabb546aa0e3283f35b686f4f3b2ff47.pdf

Anthropic. (2025a). *Claude 4: Pioneering the Next Wave of AI with Advanced Capabilities*. Retrieved from Anthropic News

Askell, A., Bai, Y., Chen, A., Drain, D., Ganguli, D., Henighan, T., Jones, A., Joseph, N., Kaplan, J., Kendall, J., Kravec, J., Lopyrev, K., McCandlish, S., Olsson, C., Olah, C., Perez, E., Ringer, S., Johnston, T., Hatfield-Dodds, Z.,... Clark, J. (2021). *A general language assistant as a laboratory for alignment*. arXiv. https://arxiv.org/abs/2112.00861

Awad, E., Dsouza, S., Kim, R., Schulz, J., Henrich, J., Shariff, A., Bonnefon, J.-F., & Rahwan, I. (2018). The Moral Machine experiment. *Nature*, *563*(7729), 59–64. https://doi.org/10.1038/s41586-018-0637-6

Bai, Y., Kadavath, S., Kundu, S., Askell, A., Kernion, J., Jones, A., Chen, A., Goldie, A., Mirhoseini, A., McKinnon, C., Chen, C., Olsson, C., Olah, C., Hernandez, D., Drain, D., Ganguli, D., Li, D., Tran-Johnson, E., Perez, E., … Kaplan, J. (2022). *Constitutional AI: Harmlessness from AI feedback*. arXiv. https://arxiv.org/abs/2212.08073

Beauchamp, T. L., & Childress, J. F. (1979). *Principles of biomedical ethics*. Oxford University Press.

Bostrom, N. (2014). *Superintelligence: Paths, dangers, strategies*. Oxford University Press.

Brandt, R. B. (1979). *A theory of the good and the right*. Clarendon Press.

Brophy, M. E. (2009). *Moral intuitions in reflective equilibrium: Applying scientific methodology to ethics* (Doctoral dissertation). University of Minnesota. Retrieved from https://www.academia.edu/20939750/Moral_intuitions_in_reflective_equilibrium_applying_scientific_methodology_to_ethics

Christiano, P. F., Leike, J., Brown, T. B., Martic, M., Legg, S., & Amodei, D. (2017). Deep reinforcement learning from human preferences. In *Advances in Neural Information Processing Systems 30 (NIPS 2017)* (pp. 4299–4307). Curran Associates, Inc.

Dancy, J. (2013). *Moral Particularism*. In E. N. Zalta (Ed.), *The Stanford Encyclopedia of Philosophy* (Fall 2013 Edition). Retrieved from Stanford Encyclopedia of Philosophy

Daniels, N. (1979). Wide reflective equilibrium and theory acceptance in ethics. *Journal of Philosophy*, *76*(5), 256–282. https://doi.org/10.2307/2025881

Ethayarajh, K., Xu, W., Muennighoff, N., Jurafsky, D., & Kiela, D. (2024). *KTO: Model alignment as prospect theoretic optimization* [Preprint]. arXiv. https://doi.org/10.48550/arXiv.2402.01306Wikipedia+10

Ferreira, P., Aziz, W., & Titov, I. (2025). *Truthful or fabricated? Using causal attribution to mitigate reward hacking in explanations*. arXiv. https://arxiv.org/abs/2504.05294

Ganguli, D., Lovitt, L., Kernion, J., Askell, A., Bai, Y., Kadavath, S.,... & Amodei, D. (2023). *Collective constitutional AI: Aligning a language model with public input*. Anthropic. https://www.anthropic.com/index/collective-constitutional-ai-aligning-a-language-model-with-public-input

Google. (2018). *Artificial intelligence at Google: Our principles*. Google AI. Retrieved from https://ai.google/responsibility/principles/

Greene, J. D. (2014). *Moral tribes: Emotion, reason, and the gap between us and them*. Penguin Press.

Greene, J. D., Sommerville, R. B., Nystrom, L. E., Darley, J. M., & Cohen, J. D. (2001). An fMRI investigation of emotional engagement in moral judgment. *Science*, *293*(5537), 2105–2108. https://doi.org/10.1126/science.1062872

Haidt, J. (2012). *The righteous mind: Why good people are divided by politics and religion*. Pantheon Books.

Haidt, J., Graham, J., & Joseph, C. (2012). Moral foundations theory: The pragmatic validity of moral pluralism.

Hare, R. M. (1981). *Moral thinking: Its levels, method, and point*. Oxford University Press.

Harman, G. (2004). Reflective equilibrium. In S. P. Stich & T. A. Warfield (Eds.), *The Blackwell guide to philosophy of mind* (pp. 13-28). Blackwell Publishing.

Haugeland, J. (1985). *Artificial Intelligence: The Very Idea*. MIT Press.

Hong, Y., et al. (2024). *ORPO: Monolithic Preference Optimization without Reference Model*. Retrieved from arXiv

Kahneman, D. (2011). *Thinking, fast and slow*. Farrar, Straus and Giroux.

Keswani, V., Conitzer, V., Sinnott-Armstrong, W., Nguyen, B. K., Heidari, H., & Schaich Borg, J. (2025). *Can AI model the complexities of human moral decision-making? A qualitative study of kidney allocation decisions*. In *Proceedings of the 2025 CHI Conference on Human Factors in Computing Systems* (CHI '25). Association for Computing Machinery. https://doi.org/10.1145/3706598.3714167

Korsgaard, C. M. (1996). *The sources of normativity*. Cambridge University Press.

Lakatos, I. (1970). Falsification and the methodology of scientific research programmes. In I. Lakatos & A. Musgrave (Eds.), *Criticism and the growth of knowledge* (pp. 91–196). Cambridge University Press.

Microsoft. (n.d.). *Responsible AI*. Microsoft. Retrieved from https://www.microsoft.com/en-us/ai/responsible-ai

Ord, T. (2020). *The precipice: Existential risk and the future of humanity*. Hachette Books.

Ouyang, L., Wu, J., Jiang, X., Almeida, D., Wainwright, C. L., Mishkin, P., Zhang, C., Agarwal, S., Slama, K., Ray, A., Schulman, J., Hilton, J., Kelton, F., Miller, L., Simens, M., Askell, A., Welinder, P., Christiano, P., Leike, J., & Lowe, R. (2022). *Training language models to follow instructions with human feedback*. arXiv. https://arxiv.org/abs/2203.02155 (NeurIPS 2022).

Rae, J., et al. (2021). *Scaling Language Models: Methods, Analysis & Insights from Training Gopher*. DeepMind. arXiv. https://arxiv.org/abs/2112.11446.

Rafailov, R., et al. (2023). *Direct Preference Optimization: Your Language Model is Secretly a Reward Model*. Retrieved from arXiv

Rawls, J. (1971). *A theory of justice*. The Belknap Press of Harvard University Press.

Rawls, J. (1993). *Political liberalism*. Columbia University Press.

Sayre-McCord, G. (2021). *Moral Epistemology*. In E. N. Zalta (Ed.), *The Stanford Encyclopedia of Philosophy* (Fall 2021 Edition). Stanford University. https://plato.stanford.edu/entries/moral-epistemology/

Scanlon, T. M. (1998). *What we owe to each other*. The Belknap Press of Harvard University Press.

Scanlon, T. M. (2003). Rawls on justification. In S. Freeman (Ed.), *The Cambridge companion to Rawls* (pp. 139–167). Cambridge University Press.

Singer, P. (2005). Ethics and intuitions. *The Journal of Ethics*, *9*(3-4), 331–352. https://doi.org/10.1007/s10892-005-3508-y

United Nations. (1948). *Universal Declaration of Human Rights* (General Assembly resolution 217 A (III)).

Williamson, T. (2007). *The philosophy of philosophy*. Blackwell Publishing.

Zhao, H., Chen, H., Yang, F., Liu, N., Deng, H., Cai, H., Wang, S., Yin, D., & Du, M. (2023). Explainability for large language models: A survey. *ACM Computing Surveys*. https://arxiv.org/abs/2309.01029